\documentclass{PoS}

\title{The Camera Calibration Strategy of the Cherenkov Telescope Array}

\ShortTitle{CTA Camera Calibration Strategy}

\author{\speaker{M.~K.~Daniel}\\
        Department of Physics, University of Liverpool, Liverpool, L69 7ZE. UK.\\
        E-mail: \email{michael.daniel@liverpool.ac.uk}}

\author{M.~Gaug\\
        Unitat de F\`{\i}sica de les Radiacions, Departament de F\`{\i}sica, and CERES-IEEC, Universitat Aut\`{o}noma de Barcelona, E-08193 Bellaterra, 
Spain \\
        E-mail: \email{markus.gaug@uab.cat}}

\author{P.~Majumdar\\
        Saha Institute of Nuclear Physics, 1/AF Bidhannagar, Sector-I Kolkata-700064, India\\
        E-mail: \email{pratik.majumdar@saha.ac.in}}

\author{for the CTA Consortium\\
        https://www.cta-observatory.org/}

\abstract{
The Cherenkov Telescope Array (CTA) will be the next generation 
ground based observatory in very high energy gamma ray astronomy. 
The facility will achieve a wide energy coverage, starting from a 
threshold of a few tens of GeV up to hundreds of TeV by utilising 
several classes of telescopes, each optimised for different regions 
of the gamma-ray spectrum. The required energy resolution of better 
than 10-15\% over most of the energy range and a goal of 5\% 
systematic uncertainty on the measurement of the Cherenkov light 
intensity at the position of each telescope means that a very precise 
evaluation of the entire system will need to be made. The composite 
nature of the array means multiple camera technologies will be 
employed so multiple calibration systems and procedures will be 
necessary to meet the performance requirements. Additional 
constraints will come from the need to minimise observing time 
losses and that the observatory is foreseen to operate for tens of 
years, so both short and long term systematic changes in performance 
will need to be investigated and monitored. This contribution 
summarises the recommended camera calibration strategy of CTA based 
on the experience with current IACTs.
}

\FullConference{The 34th International Cosmic Ray Conference,\\
		30 July- 6 August, 2015\\
		The Hague, The Netherlands}

\begin{document}

\section{Introduction}

The camera of an imaging atmospheric Cherenkov telescope (IACT) consists of an array of pixels and their readout system. It begins with the photodetector that converts a Cherenkov photon to a  photoelectron (p.e.) and amplifies this to a measurable current or voltage, which is in turn digitised and stored as digital counts (d.c.). The precision of the camera calibration will ultimately determine how accurately we can later reconvert the d.c. and use them to estimate the Cherenkov light intensity that arrived at the camera focal plane. Most of these calibration parameters will be \emph{characterised} in the laboratory during the commissioning stage of the camera~\cite{CTF}, but these initial values can vary due to ambient effects (e.g. temperature or night-sky background illumination level) or through ageing in the harsh operating environment. Here we refer to \emph{calibration} as the on-site measurement and update of the lab. \emph{characterisation} values, with measurements to be taken at intervals specific to the expected timescale of variation in the respective coefficient.

The Cherenkov Telescope Array (CTA)\footnote{www.cta-observatory.org} \cite{CTA} will be the next generation IACT facility, providing the community with an open-access observatory for the observation of gamma rays with energies from a few tens of GeV to hundreds of TeV with unprecedented sensitivity and similar improvement in angular resolution and energy resolution. To achieve this large dynamic range in energy coverage the array will be comprised of multiple telescope sizes: a few 23\,m diameter large size telescopes (LSTs) provide a low energy threshold; several 12\,m diameter medium size telescopes (MSTs) provide an order of magnitude improvement in flux sensitivity in the 0.1 to 10\,TeV region; and many 4\,m diameter small sized telescopes (SSTs) will examine the lowest flux, highest energy events ($E \geq 10$\,TeV). To meet the science case requirements the systematic uncertainty of the energy of a photon candidate (at energies above 50 GeV) must be $<15\%$ and the energy resolution must be $<30\%$ at 50\,GeV and $<10\%$ above 1\,TeV; with the systematic uncertainty on the overall energy scale to not exceed 10\%.  The linear response of the CTA cameras must be recoverable to $\leq 5\%$ up to 1000 p.e. per pixel. The MSTs and SSTs will also comprise of both single (-1M) and dual mirror (-2M) optics, leading to some very different camera configurations being used in the array.

There are currently seven camera prototypes proposed for use in CTA, employing a wide variety of photodetector and data acquisition technologies:
\begin{description}
\item[LST camera] The LSTs will have AC coupled photomultipliers sampled by DRS4 Switched Capacitor Arrays \cite{LSTDACQ}, this camera is optimised to lower the energy threshold by triggering a single telescope deep in to the night sky background (NSB) regime. A large dynamic range is achieved through the use of two gain channels.
\item[FlashCam] A camera design for the MST \cite{FlashCam} with DC coupled photomultipliers sampled by FADC and achieving a large dynamic range with non-linear gain pre-amplifiers of two distinct response regimes: a linear and a logarithmic region for small and large amplitudes respectively and a smooth transition between these regimes.
\item[NectarCam] A camera design for the MST with AC coupled photomultipliers sampled by the NECTAr chip~\cite{NectarCam}. It also achieves its dynamic range through the use of two gain channels.
\item[SCT camera] A camera for the dual-mirror Schwarzchild Couder Telescope (SCT) utilising Silicon Photomultiplier (SiPM) readout by a TARGET ASIC~\cite{TARGET} that samples the waveform~\cite{SCTDACQ}.
\item[SST-1M camera] An adaptation of the FlashCam for the SST-1M, in this case DC coupled silicon photomultipliers (SiPM) are employed on the focal plane \cite{DigiCam}.
\item[GCT camera] A camera for a SST-2M. The photodetector is either Multi-Anode photomultiplier (MAPM) or SiPM that has the waveform sampled by a TARGET ASIC \cite{GCT}. The camera saturates at about $\sim$600\,p.e. but signal fitting (e.g. Time Over Threshold) can recover the signal to $\leq$5\% bias up to $\sim$1000\,p.e. 
\item[ASTRI] A camera for a SST-2M. AC coupled SiPM photosensors are read out by CITIROC which stores the peak of the electronically shaped signal in a pre-defined readout window width \cite{ASTRI}. The dynamic range is achieved by having a low gain channel and a high gain channel.
\end{description}
In light of this complexity it is necessary to have a clearly established set of guidelines for the calibration of the camera data to ensure that the CTA performance requirements are met, whilst simultaneously minimising the complexity of the data analysis chain in order to simplify the software maintenance costs and make for easier upgrades. These guidelines will also assure the minimisation in the complexity and fulfill the scientific requirements in the real time analysis~\cite{RealTimeAnalysis} as well as offline in the data centre.

\section{Measurement of light intensity}
\label{sec:intensity}

The response of a pixel ($i$) of solid angle $\Omega$ to a light of intensity $I$ (photons per unit area and solid angle) is given by
\begin{equation}
V_i (I) =  kI + p_i = I \cdot A_\mathrm{eff} \cdot R_\mathrm{Mir} \cdot \Omega \cdot  T_{F,i} \cdot \eta_{Q,i} \cdot \eta_{c,i} \cdot e \cdot G_{S,i}\cdot G_{el,i} + p_i
\label{eqn:PixelPEResponse}
\end{equation}
where $V(I)$ is the output in digital counts, possibly after some waveform processing and/or non-linearity corrections; 
$A_\mathrm{eff}$ the effective mirror area after shadowing; 
$R_\mathrm{Mir}$ the focused mirror reflectivity; 
$T_F$ the transmission efficiency through the optical system, e.g. shadowing, lightcones, window, etc;
$\eta_Q$ the sensor quantum efficiency; 
$\eta_c$ the p.e. collection efficiency; 
$e$ the electron charge; 
$G_S$ the gain in the sensor; 
$G_\mathrm{el}$ the electronics conversion factor in terms of d.c. per p.e.; and $p$ a baseline offset. For simplicity, we have omitted here the possible dependence of the factors on the angle of incidence, and assumed that quantities vary independently of the photon wavelength. In reality the dual mirror telescope designs will have photons striking the focal plane at much larger angles of incidence than the single mirror designs and also at different angles to any telescope mounted calibration light sources, which could introduce systematic differences of up to $\sim 30\%$ if not properly accounted for. The degradation of the transmission and quantum efficiency coefficients is unlikely to be achromatic either and must again be carefully accounted for, especially if there is any bias between the Cherenkov light spectrum from the air shower and that of the calibration light source (e.g. local muons).

For CTA, the systematic error on the measurement of the absolute intensity (i.e. photons per square metre) of the Cherenkov light (post-calibration) at the position of each telescope must be $<8$\%, with a goal of $\leq 5\%$. 
The uncertainty in the estimate of the Cherenkov light intensity at each
telescope is a combination of the uncertainties in the optical path efficiency, 
the Cherenkov photon detection efficiency at the focal plane and the statistical
and systematic uncertainties in the photon conversion, photo-electron 
amplification and digitisation in the camera. Assuming a camera averaged
response from eq.~\ref{eqn:PixelPEResponse} and 
assuming there are no dominant hotspots to bias the result we see that the 
error ($\delta S$) in the Cherenkov image size ($S$, equivalent to the intensity) depends on
\begin{equation}
 \left ( \frac{ \delta S} {S} \right )^{2} = 
 \left ( \frac{ \delta V} {V} \right )^{2} + 
 \left ( \frac{ \delta p} {S} \right )^{2} + 
 \left ( \frac{ \delta \textit{FF}} {\textit{FF}} \right )^{2} + 
 \left ( \frac{ \delta T} {T} \right )^{2} +
 \left ( \frac{ \delta \textit{MC}} {\textit{MC}} \right )^{2} 
\end{equation}
where $\delta V$ is the uncertainty in the electronics signal, which contains the photon conversion efficiency, and uncertainties in knowledge of the precise amplification  and digitization; $\delta p$ is the uncertainty from the pixel baseline; $\delta$\textit{FF} is the uncertainty in the camera flat-fielding; $\delta T$ is the uncertainty in the light transmission; and  $\delta$\textit{MC} is the uncertainty introduced by approximations used in the Monte Carlo simulation. If we divide the error budget equally between the photon transmission before reaching the camera (i.e. mirrors, lightguides, window and shadowing) and from the conversion to p.e. onwards (i.e. photodetection, amplification and digitisation) we can reach the goal requirement allowing 1\% uncertainty in the digitisation, 0.2\,p.e. uncertainty in the pixel baseline measurement and 2\% uncertainty in the flat-fielding of the camera response. Some, like the uncertainty in the pedestal baseline, have already been shown to be achievable on current generation instruments; others, like measuring the uncertainty on the transmission, are a step further in the calibration demands compared to current instruments (due to the additional need to control the wavelength and angular dependence aspects); others still, like flat-fielding the curved focal plane of the dual mirror telescope systems from short illumination distances have presented challenges never before faced in this field and so novel techniques must be thought of (e.g.~\cite{CHECCal, Illuminator}). If the chromatic biases between Cherenkov light from muons and from air showers can be kept to a minimum then an end-to-end calibration using muons rings with a precision of $\leq 5\%$ can be met, but resolving the individual contributions can still be challenging.

\begin{figure}[htbp]
\begin{center}
\resizebox{0.6\textwidth}{!}{\includegraphics{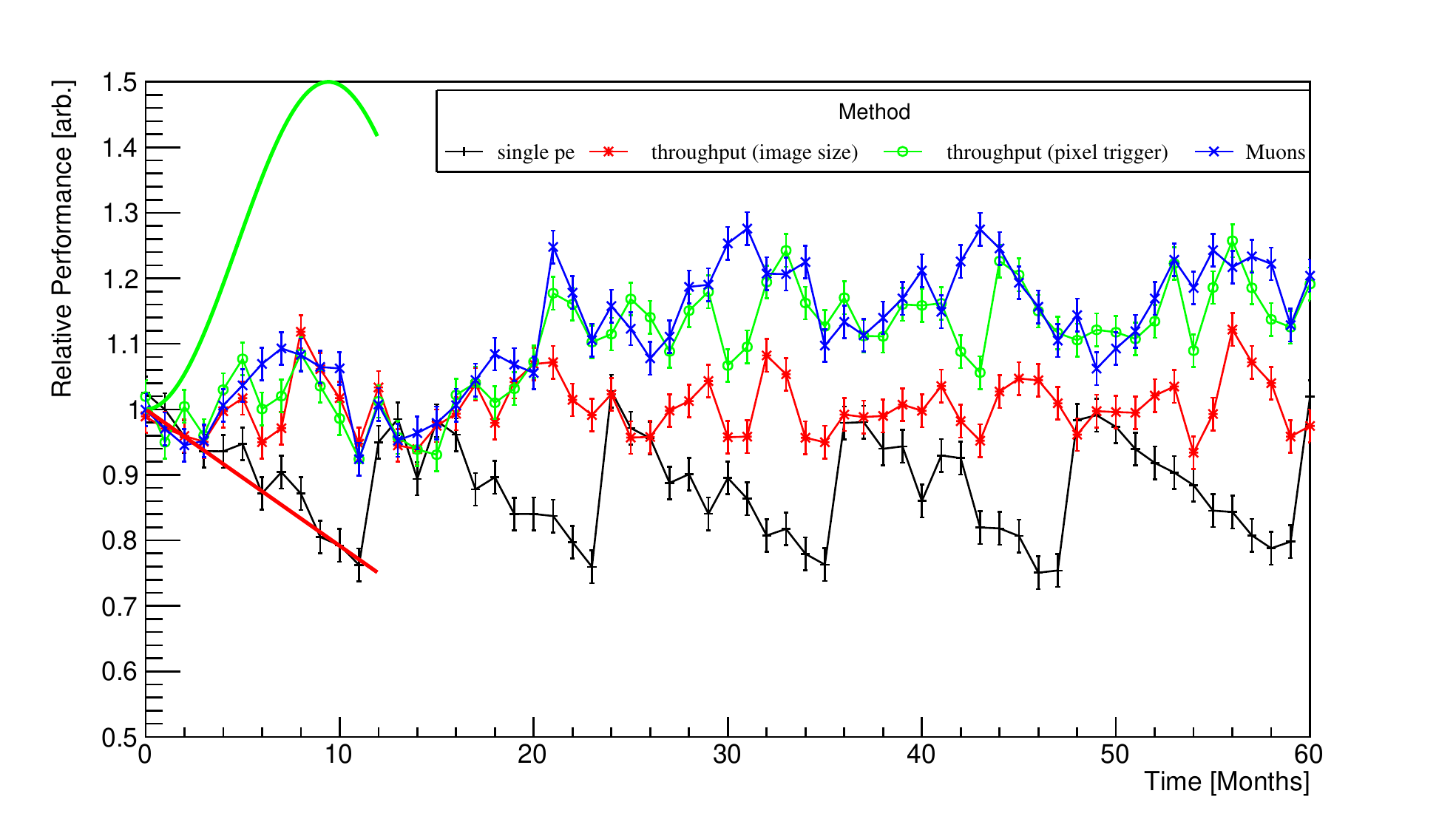}}
\caption[]
        {Mock up of a telescope's relative performance over a 5 year period as measured by four different methods: 
	  single p.e. (black cross); 
	  cosmic-ray throughput measured by image size (red star); 
	  cosmic-ray throughput measured by 2nd brightest pixel (green circle);
	  muon rings (blue cross). 
	 Degradation is exaggerated to clearly see the differences in the figure. The error bars are for a 2.5\% random statistical error that has been included for the data points. Systematic effects including PMT ageing (red line), seasonal variation in Cherenkov light yield (green line) with an additional aerosol layer in months 6 \& 7 of a year and periodic changes including a yearly increase in the gain and a refocusing of the camera at month 20 are also plotted.}
\label{fig:RelPerf}
\end{center} 
\end{figure}

Tracking where issues in the systematic uncertainties may arise is best achieved when multiple, or redundant, methods are at ones disposal.  Figure~\ref{fig:RelPerf} shows a mock data set for the relative efficiency of a generic camera over a 5 year period as measured by the multiple methods of calculating the p.e./d.c. value from single photoelectron runs and from using muon rings, and the relative measurements of cosmic-ray throughput \cite{Lebohec03} based on image size and from triggering pixels only. Each method is sensitive to different levels of systematic effects that change the expected number of digital counts received from the camera and in turn to how the system is simulated to produce lookup tables for data analysis. To exaggerate the effects for display purposes the camera is simulated with a 20\% drop in pixel efficiency per year through ageing (compensated for by raising the voltages once per year); a 25\% change in the Cherenkov light yield due to seasonal variations in atmospheric density and a 10\% drop in transmission due to an enhanced high altitude aerosol layer (such as from calima or biomass burning) in month 6 and 7 of a year; and a 15\% change in the light focused onto a pixel occuring 20 months after the start of observatory operations, e.g. as might be expected if the telescope were refocused between infinity and the altitude of shower maximum. Only at very specific times do all the methods give consistent results. The single photoelectron measurement will only be sensitive to changes in the camera specific hardware, whereas the other end-to-end measurements are also susceptible to the systematics of the atmosphere and the telescope structure. The seasonal changes in the atmosphere at certain times can counteract the camera hardware ageing, at others enhance them. The change in focus affects measurements that use the per pixel amount of light, such as muon rings and triggering pixels, but not ones that use all of the image light, hence the splitting between the different throughput measurements at month 20. Where achievable we recommend a primary method for the calculation of calibration parameters and secondary (or tertiary) methods that can be used as cross-checks on a less frequent basis, or in the event of failure of hardware for the primary method. 
An additional consideration to take into account is that a minimal amount of time must be taken from the observing schedule to deal with calibration issues in dark time. As such, any calibration that can be performed simultaneously with the acquisition of observational data must do so, this will also have the added benefit of giving calibration information that is relevant for the current ambient conditions at the time of data taking.

\section{Calibration hardware and procedures}

Hardware, procedures, methods and interfaces for the following calibration must be in place, though not every one will be relevant to all camera designs. They fall in to the categories of regular monitoring ($\geq$ once per night, e.g.~\cite{SST1MCamCal}) and long term monitoring (at least once per year, that check the coefficients measured from the camera characterisation stage~\cite{CTF}).

\subsection{Regular Monitoring: on telescope systems}
\label{sec:flasher}

\begin{description}
\item[Pixel baseline and its fluctuations] Should be determined under actual observing conditions in order to account for potential baseline shifts due to background illumination (depending e.g. on whether electronics is AC or DC coupled) and ambient environmental effects (e.g. temperature).  
Various well-proven techniques exist to determine pixel baseline, e.g.: 
random-triggered events can be injected into the data stream (this is the preferred scheme for CTA); alternatively, baselines can be determined from blank regions of shower images; or in the case of waveform readout, baselines can be determined from samples preceding the triggered signal, provided a sufficiently wide readout window is used. The electronics baseline (i.e. independent of NSB contribution) can be determined by force triggering the camera with the camera lid closed.

\item[Relative pixel response to incident light (intensity flat fielding)]
Flat-fielding of the camera can be achieved by two methods: illumination of the camera with a light pulser with a light front of known intensity variation (preferably uniform); or the pixel response for muon rings. The light pulser will usually be the simplest and most effective solution. 

\item[Relative pixel response to incident light (timing flat fielding)]
The timing offsets between pixels should be identified to sufficient precision using the intensity flat-fielding events.

\item[Effective number of photoelectrons / excess noise ]
Several techniques can be employed to determine the gain of a photosensor: the fluctuation in the response to large calibration light pulses (see e.g. \cite{Hanna2010,mirzoyan1997}), which is the easiest scheme to implement on a nightly basis whilst minimising the impact to the observation schedule by using the flat-fielding events; 
exposure of the camera pulsed light at low level, under suitably dark conditions (i.e. using a shelter or camera mask), such that the single photoelectron signal can be directly determined, which can be used as an accurate cross-check on a less frequent basis, or comes for free with SiPM based cameras; exposure with night-sky background, using timing information to select pixels which have a signal likely related to one or few night-sky photoelectrons.
\end{description}

On telescope light flasher systems can be simple, and serve primarily for flat-fielding. LED systems with pulse width and spectrum roughly matched to the Cherenkov spectrum seem most promising, following e.g. the concept of~\cite{CHECCal}, though the exact IACT requirements may still require the use of a laser based system, e.g.~\cite{LSTCamCal, FlashCam}. They should provide a known illumination light front shape  of the camera with a fixed pulse height of $>30$\,p.e. and preferably 100\,p.e.. Strict long-term stability of the flash amplitude is not required, short-term stability (minute-scale) must be good compared to the statistical variation of the pixel signals. It should be possible to trigger the flasher during a run to inject calibration events without significantly affecting observations.

\subsection{Long Term Monitoring: high-performance light flasher systems}
\label{sec:illuminator}
A high-performance light flasher system should be available to characterize telescopes at longer intervals (e.g. once per year). The flasher system can either be placed on the IACT, or -- possibly simpler but requiring larger light intensity -- a movable system that is placed at a set distance from the telescope (e.g.~\cite{Illuminator}). In the latter case, it should reasonably uniformly illuminate the whole mirror surface.
The system should ideally provide the following features:
\begin{itemize}
\item Combination of pulsed and DC light
\item DC light with selectable wavelength over the range of sensor sensitivity, and adjustable intensity up to roughly full-moon equivalent. Absolute calibration of the DC light intensity is not required.
\item Pulsed light with selectable wavelength over the range of sensor sensitivity, and adjustable intensity up to saturation of the sensors or electronics chain. Pulse length should match Cherenkov pulse length. Relative light level should be calibrated with a precision of better than 3\%, e.g. using a precision optical attenuator. Absolute light level should be calibrated with a precision of 5\%, e.g. using calibrated photodiodes at some distance.
\item Built-in battery power and wireless control would increase flexibility.
\end{itemize}
The system will then be used to measure
\begin{description}
\item[Pixel linearity] To determine pixel response at several intensity levels up to and beyond 1000\,p.e., particularly for dual gain channel or non-linear amplification systems.
\item[Pixel spectral response] Spectral response needs to be measured in the lab and occasionally be verified in the field. Spectral response must be known sufficiently well that the calibration reference light source yield relative to the air shower light yield and night-sky background response is known with a precision at a level of $\sim 1-2$\% -- this is especially true for calibration methods using the Cherenkov light emitted by muons. A calibration system providing four colours ranging from UV to red should be at least available once per year.
\item[Pixel response for different light incidence angles] An appropriate solution would be to have a high-performance angular response calibration system, consisting of a single~pe illumination source that can illuminate at a range of angles from face-on to the maximum expected ray angle (up to 70$^\circ$ for the dual mirror telescope systems), available in the on-site lab to calibrate a representative number of pixels off-telescope, and a system placing a remote light source at about twice the focal distance, with the telescope parked and the reference light front thus following a similar optical path via the mirrors to the focal plane; however, shadowing issues and reproducibility of the method must be demonstrated beforehand. 
\item[Integration Window]  Special calibration runs with the extraction window at maximum can be compared to runs with the standard shorter operating window. Flat-fielding events with a similar FWHM to Cherenkov events can be used to determine an appropriate window size relatively quickly, but Cherenkov images will have a larger time structure across the image which can result in signal loss. It is recommended to determine these losses through specific runs with the telescopes pointing at zenith and comparing the image size distributions as a function of the reconstructed impact distance.
\end{description}

\section{Summary}
The ambitious science requirements and long operational life span of CTA, combined with the increased complexity of operating multiple telescope types, means that clearly defined calibration procedures in the IACT camera calibration need to be in place if systematic uncertainties are to be kept to a minimum. The camera calibration can be split according to whether a coefficient needs to be regularly or infrequently measured: the former preferably done with events that are taken in conjunction with science observations; the latter that can be done with more specialised hardware in dedicated campaigns.

We gratefully acknowledge support from the agencies and organizations listed 
under Funding Agencies at this website: http://www.cta-observatory.org/.

\end{document}